\documentstyle [12pt,axodraw] {article}

\catcode`@=12
\begin{document}
\thispagestyle{empty}
\begin{flushright} UCRHEP-T222\\April 1998\
\end{flushright}
\vspace{0.5in}
\begin{center}
{\Large	\bf Pathways to Naturally Small Neutrino Masses\\}
\vspace{1.5in}
{\bf Ernest Ma\\}
\vspace{0.3in}
{\sl Department of Physics, University of California\\}
{\sl Riverside, California 92521, USA\\}
\vspace{0.1in}
\vspace{1.5in}
\end{center}
\begin{abstract}\
In the minimal standard electroweak gauge model, there is an effective 
dimension-five operator which generates neutrino masses, and it has only 
three tree-level realizations.  One is the canonical seesaw mechanism with 
a right-handed neutrino.  Another is having a heavy Higgs triplet as recently 
proposed.  The third is to have a heavy Majorana fermion triplet, an example 
of which is presented here in the context of supersymmetric $SU(5)$ grand 
unification.  The three generic one-loop realizations of this operator are 
also discussed.
\end{abstract}

\newpage
\baselineskip 24pt

In the minimal standard gauge model of quarks and leptons, each of the three 
known neutrinos $(\nu_e, \nu_\mu, \nu_\tau)$ appears only as a member of a 
left-handed $SU(2)$ lepton doublet
\begin{equation}
\psi_i = (\nu_i, l_i)_L,
\end{equation}
and the Higgs sector contains only one scalar doublet
\begin{equation}
\Phi = (\phi^+, \phi^0).
\end{equation}
As a result, neutrinos are massless in this model.  Experimentally there 
is now a host of evidence for neutrino oscillations, and that is most 
naturally explained if neutrinos are massive and mix with one another. 
Theoretically there is no compelling reason for massless neutrinos and 
any extension beyond the minimal standard model often allows them to be 
massive.  There exists already a vast literature on specific models of 
neutrino mass and mixing.  

In this paper I make the following simple observation.  In the minimal 
standard electroweak gauge model, there is an effective dimension-five 
operator which generates Majorana neutrino masses, to wit
\begin{equation}
\Lambda^{-1} \phi^0 \phi^0 \nu_i \nu_j,
\end{equation}
where $\Lambda$ is a large effective mass.  All models of neutrino mass 
and mixing (which have the same light particle content as the minimal 
standard model) can be summarized by this operator.  Different models 
(among them the well-known seesaw model\cite{1}) are merely different 
realizations of this operator.  In the following I will show that it has 
only three tree-level realizations, all of which are conceptually just as 
simple.  I will also discuss its many possible one-loop realizations, 
encompassing thus most previous work on radiative neutrino masses.

To obtain the effective operator (3) at tree level, using only renormalizable 
interactions, it is clear that there are only three ways.
\begin{center}
\begin{tabular} {rl}
(I) & $\psi_i$ and $\Phi$ form a fermion singlet, \\
(II) & $\psi_i$ and $\psi_j$ form a scalar triplet, \\
(III) & $\psi_i$ and $\Phi$ form a fermion triplet.
\end{tabular}
\end{center}
Note that the singlet combination of $\psi_i$ and $\psi_j$ is $\nu_i l_j - 
l_i \nu_j$ which does not generate (3).  In each case, the complete 
gauge-invariant effective operator is actually the same, but how it is 
written reveals its possible origin:
\begin{eqnarray}
{\rm (I)} && \Lambda^{-1} (\phi^0 \nu_i - \phi^+ l_i)(\phi^0 \nu_j - \phi^+ 
l_j), \\ {\rm (II)} && \Lambda^{-1} [\phi^0 \phi^0 \nu_i \nu_j - \phi^+ 
\phi^0 (\nu_i l_j + l_i \nu_j) + \phi^+ \phi^+ l_i l_j], \\ 
{\rm (III)} && \Lambda^{-1} [(\phi^0 \nu_i + \phi^+ l_i)(\phi^0 \nu_j + 
\phi^+ l_j) - 2 \phi^+ \nu_i \phi^0 l_j - 2 \phi^0 l_i \phi^+ \nu_j].
\end{eqnarray}

(I) ~ The intermediate heavy particle in this case is clearly a 
fermion singlet as shown in Fig.~1.  Call it $N$ and let its mass be $M$ 
and its coupling to $\nu_i$ be $f_i$, then $\Lambda^{-1} = f_i f_j/2M$.  
As $\phi^0$ acquires a nonzero $vev$ [vacuum expectation value $\langle 
\phi^0 \rangle = v$], we can identify $f_i v$ as a Dirac mass $m_i$ linking 
$\nu_i$ to $N$ and the neutrino mass matrix is 
simply $-m_i m_j/M$.  This is of course just the well-known seesaw mechanism, 
with $N$ identified as the right-handed neutrino with a large Majorana mass.  
Note that with one such singlet, only one linear combination of $\nu_i$ 
gets a tree-level mass.  Hence the usual scenario requires three 
$N$'s.  This mechanism of generating naturally small 
neutrino masses dominates the literature, but as I show below, the 
two other alternatives are conceptually just as simple.

(II) ~ What is needed here is a heavy scalar triplet $\xi = (\xi^{++}, 
\xi^+, \xi^0)$ as shown in Fig.~2.  If its mass is $M$ and its coupling to 
$\nu_i \nu_j$ and $\phi^0 \phi^0$ are $f_{ij}$ and $\mu$ respectively, then 
$\Lambda^{-1} = f_{ij} \mu/M^2$ and the neutrino mass matrix is given by
\begin{equation}
\left( {\cal M}_\nu \right)_{ij} = {-2 f_{ij} \mu v^2 \over M^2}.
\end{equation}
Note that only one $\xi$ is required for all neutrinos to become massive. 
This is a simple mechanism which does not require right-handed neutrinos, 
and is indistinguishable from (I) as far as the low-energy limit of the 
theory is concerned.  As already discussed recently\cite{2}, another way of 
understanding the above is to consider the $vev$ of $\xi$.  Although 
$\xi$ is very heavy, it acquires a tiny $vev$ given by $u = -\mu 
v^2/M^2$, hence the neutrino mass matrix is equal to $2 f_{ij} u$ as 
expected from the direct coupling of $\xi$ to $\nu_i \nu_j$.  The idea 
that a heavy Higgs scalar could have a naturally small $vev$ was known but 
not widely appreciated and this mechanism has largely been neglected.

(III) ~ We replace $N$ of (I) here with a heavy Majorana fermion 
triplet $(\Sigma^+, \Sigma^0, \Sigma^-)$.  Again a seesaw mass is 
obtained\cite{3} and there is no low-energy distinction between this and the 
other two mechanisms.  However, each has its own unique implications about 
physics beyond the standard model.  In (I), the addition of three $N$'s argues 
favorably for the efficacy of $SO(10)$ instead of $SU(5)$ as a suitable 
unifying symmetry, whereas in (II) and (III), $SU(5)$ by itself 
is sufficient.  The {\bf 15} representation of $SU(5)$ would contain $\xi$, 
whereas the {\bf 24} representation would contain both $N$ and $\Sigma$.

As an example, consider a supersymmetric $SU(5)$ model of grand 
unification\cite{4}. The breaking of $SU(5)$ to the standard $SU(3) \times 
SU(2) \times U(1)$ gauge group is accomplished using the {\bf 24} 
supermultiplet,
\begin{equation}
{\bf 24} = (1,1,0) + (8,1,0) + (1,3,0) + (3,2,-5/6) + (3^*,2,5/6),
\end{equation}
where the scalar component of (1,1,0) acquires a large $vev$.  The fermionic 
components of (1,1,0) and (1,3,0) are exactly $N$ and $\Sigma$ of (I) 
and (III).  However, a $\nu_i \phi^0 N$ coupling is not desirable 
because the scalar partner of $N$ has a large $vev$ and $\nu_i$ must then 
combine with the fermion partner of $\phi^0$ to form a superheavy Dirac 
particle.  On the other hand, since the scalar partner of $\Sigma^0$ has no 
$vev$, a $\nu_i \phi^0 \Sigma^0$ coupling is permissible in principle.

Now $\psi_i$ belongs to the {\bf 5}$^*$ representation 
and there are two scalar doublets $\Phi_1 = (\phi_1^0, \phi_1^-)$ and 
$\Phi_2 = (\phi_2^+, \phi_2^0)$ belonging to the {\bf 5}$^*$ and {\bf 5} 
representations respectively.  With only one {\bf 5} and one {\bf 24}, there 
can be only one {\bf 5}$^*$ which appears in the singlet decomposition of 
{\bf 5}$^* \times$ {\bf 5} $\times$ {\bf 24}.  This {\bf 5}$^*$ is defined 
to be the one containing $\Phi_1$.  Hence $\psi_i$ does not couple to 
$N$ or $\Sigma$ of Eq.~(8).  To obtain a neutrino mass, we need another 
{\bf 24} which has no $vev$.  Consider then a discrete $Z_2$ symmetry, under 
which all the quark and lepton supermultiplets are odd, and all others 
are even except for this additional {\bf 24} which is odd as well.  In 
that case, one linear combination of $\nu_i$  gets a seesaw mass from the 
$N$ and $\Sigma$ of this odd {\bf 24}.  
For $M_N \sim M_\Sigma \sim 10^{16}$ GeV and $\langle \phi_2^0 
\rangle \sim 10^2$ GeV, a neutrino mass of order $10^{-3}$ eV is then 
very natural and suitable for solar neutrino oscillations\cite{5}.

Consider now the MSSM (Minimal Supersymmetric Standard Model\cite{6}) which 
pervades the present literature on particle physics.  The neutrinos of this 
model are massless.  However, if the MSSM is the low-energy remnant of 
supersymmetric $SU(5)$, then an additional superheavy {\bf 24} naturally 
yields one massive neutrino which could explain the solar 
data.  However, to accommodate either the atmospheric data\cite{7} 
or the LSND data\cite{8} as well, we need another massive neutrino.  We now 
have the option of using any one of the above three mechanisms.  For example, 
if we would add another odd {\bf 24} with a mass of order $10^{14}$ GeV, we 
could get a neutrino mass of about 0.1 eV, which would be suitable for 
atmospheric neutrino oscillations.

The effective operator (3) may be realized also radiatively in one loop. 
There is in fact one well-known generic mechanism\cite{9} as shown in Fig.~3. 
The fermions $\omega$ and $\omega^c$ must couple to $\phi^0$, hence one 
of them has to belong to a doublet.  Without loss of generality, we choose
\begin{equation}
\omega \sim (q_3,2,q_1)
\end{equation}
under $SU(3) \times SU(2) \times U(1)$.  We then must have
\begin{equation}
\omega^c \sim (q_3^*,q_2,-q_1+{1 \over 2}),
\end{equation}
where $q_2 = 1$ or 3.  As we go around the loop, we see that
\begin{equation}
\eta \sim (q_3,2,q_1)
\end{equation}
as well, and
\begin{equation}
\chi \sim (q_3^*,q'_2,-q_1+{1 \over 2}),
\end{equation}
where $q'_2 = 1$ or 3 also.  For a given choice of $q_3$ and $q_1$, there 
are then 4 variations, corresponding to the choice of $q_2$ and $q'_2$. 
Most specific proposals for the one-loop radiative generation of Majorana 
neutrino masses are contained in the above.

The fermions $\omega$ and $\omega^c$ may in fact be the usual quark or 
lepton doublet and singlet.  For example, if we choose $q_3 = 1$, $q_1 = 1$, 
and $q_2 = q'_2 = 1$, then $\omega \sim (1,2,-1/2) \sim l_L$ and $\omega^c 
\sim (1,1,1) \sim l_L^c$.  Now $\eta \sim (1,2,-1/2)$ and $\chi \sim (1,1,1)$ 
may be arbitrary new scalar particles, in which case we have the Zee 
model\cite{10}, or supersymmetric scalar leptons $\tilde l$ and $\tilde l^c$, 
in which case we have the R-parity violating model\cite{11}.  In the latter 
case, we may also use the quarks and their supersymmetric scalar partners, 
{\it i.e.} $q_3 = 3$, $q_1 = 1/6$, and $q_2 = q'_2 = 1$.

For simplicity, both $q_2$ and $q'_2$ are usually chosen to be one, but 
$q'_2 = 3$ has also been considered\cite{12}.  The observation that the 
effective operator (3) comes from a specific model has also been 
made\cite{13}.  Here I start with (3) and show how all specific models 
are extracted from it.  This approach leads one naturally to 
another one-loop diagram which generates (3) as shown in Fig.~4.  This 
mechanism has rarely been used, and only in scenarios\cite{14} where one 
neutrino already has a tree-level mass.

The fermions $\omega$ and $\omega^c$ of Fig.~4 must combine to form an 
invariant mass, hence
\begin{equation}
\omega \sim (q_3, q_2, q_1), ~~~ \omega^c \sim (q_3^*, q_2, -q_1).
\end{equation}
As we go around the loop, we see that
\begin{equation}
\eta \sim (q_3, q'_2, q_1 + {1 \over 2}), ~~~ \chi \sim (q_3^*, q''_2, 
-q_1 + {1 \over 2}).
\end{equation}
If $q_2 = 1$, then $q'_2 = q''_2 = 2$.  Otherwise, $q'_2$ and $q''_2$ may 
be either $q_2 - 1$ or $q_2 + 1$ independently, except that $q'_2 \times 
q''_2$ must contain the triplet representation, hence $q'_2 = q''_2 = 1$ 
is not allowed.  Consider the following specific example.  Add to the 
standard model just one right-handed neutrino singlet $N$ with a Majorana 
mass, then we can set $\omega = \omega^c = N$, {\it i.e.} $q_3 = q_2 = 1$ and 
$q_1 = 0$.  In that case, both $\eta$ and $\chi$ are (1,2,1/2) doublets, so 
they can be the same extra scalar doublet we add to the standard model.  If 
we did not have the second doublet, then only one linear combination of 
$\nu_i$'s would get a tree-level seesaw mass from $N$, and the rest would 
become massive\cite{15} only at the two-loop level through the exchange of 
2 $W$ bosons.

In the above example with one $N$, one neutrino mass is obtained at tree level 
and the others are radiative.  If there is no $N$, the mechanism of Fig.~4 
can still be used to find radiative masses for all three neutrinos.  As an 
illustration, let $\omega$ and $\omega^c$ be charged fermion singlets: 
$\omega \sim (1,1,-1)$, $\omega^c \sim (1,1,1)$.  We must now have
$\eta \sim (1,2,-1/2) = (\eta^0, \eta^-)$ and $\chi \sim (1,2, 3/2) = 
(\chi^{++},\chi^+)$.  There are then the following invariant interaction 
terms:
\begin{center}
$(\nu_i \chi^+ - l_i \chi^{++}) \omega, ~~~ (\nu_j \eta^- - l_j \eta^0) 
\omega^c$, 
\end{center}
\begin{equation}
\chi^+ \eta^- \bar \phi^0 \bar \phi^0 + (\chi^{++} \eta^- + \chi^+ \eta^0) 
\bar \phi^0 \phi^- + \chi^{++} \eta^0 \phi^- \phi^-,
\end{equation}
which allow Fig.~4 to generate radiative neutrino masses as shown.  A 
trivial variation of Fig.~4 is to replace the quartic $\chi \eta \bar \phi^0 
\bar \phi^0$ coupling with two cubic couplings $\chi \bar \phi^0 \zeta$ and 
$\eta \bar \phi^0 \bar \zeta$, where $\zeta$ is an extra complex scalar 
multiplet.

Finally, consider Fig.~5 which requires only one complex scalar multiplet 
$\zeta$ but four fermion multiplets $\omega$, $\omega^c$, $\sigma$, and 
$\sigma^c$.  This mechanism is largely known\cite{9} only for generating 
masses for quarks and charged leptons.  A variation of it was applied\cite{11} 
to Majorana neutrinos in the supersymmetric R-parity violating model, but 
there the scalar neutrinos have $vev$'s, whereas the assumption here is that 
only $\phi^0$ has a $vev$.  

The fermions $\sigma$ and $\sigma^c$ of Fig.~5 
must combine to form an invariant mass.  The simplest case is to let them 
be singlets: $\sigma \sim (q_3,1,q_1)$ and $\sigma^c \sim (q_3^*,1,-q_1)$. 
Then $\omega \sim (q_3,2,q_1+1/2)$, $\omega^c \sim (q_3^*,2,-q_1+1/2)$, 
and $\zeta \sim (q_3,1,q_1)$ or $(q_3,3,q_1)$.  If we choose $q_3 = 1$ and 
$q_1 = -1$, then $\omega$ is a doublet with charges 0 and $-1$, whereas 
$\omega^c$ is a doublet with charges 2 and 1.  We see that as in the case 
of Fig.~4, exotic representations are needed for the implementation of 
this mechanism.  We can generate all other solutions systematically by 
starting with a given $SU(2)$ representation for $\sigma$ and $\sigma^c$. 
Two other variations of Fig.~5 are also possible.  We simply place 
the invariant mass to the other side of one or the other of the two 
$\phi^0$'s.

In the three tree-level realizations of the effective operator (3) for 
naturally small Majorana neutrino masses, the mass scale of the heavy 
particles involved should be very large: $10^{13}$ to $10^{16}$ GeV. Looking 
at the three identical effective interactions of Eqs.~(4) to (6), we see 
that there would be no other observable effect except for nonzero neutrino 
masses and mixing.  In the many one-loop realizations, the mass scale of 
the new particles involved depends on the model, but there is a general 
rule.  If there is a fermion doublet which does not have an invariant mass, 
then the masses of its components must come from the $vev$ of $\phi^0$, 
hence they should be of order 100 GeV and be accessible experimentally in 
the near future.  If a new particle is found, then we can use Figs.~3 to 5 
to check if the other new particles are there or not.

In conclusion, the important issue of naturally small Majorana neutrino masses 
in any extension of the standard model, which has the same light particle 
content, can be synthesized in terms of a single effective operator: 
$\Lambda^{-1} \phi^0 \phi^0 \nu_i \nu_j$.  There are three tree-level 
realizations of this operator: one is the well-known seesaw mechanism with 
a heavy singlet fermion, another is having a heavy Higgs triplet which 
naturally acquires a tiny vacuum expectation value, a third is to replace 
the singlet in the seesaw with a triplet.  The literature on neutrino 
masses is dominated by the first mechanism, but the other two are just as 
conceptually simple and would open up the options available for physics 
beyond the standard model.  There are also three one-loop realizations 
(plus variations) of this operator as shown in Figs.~3 to 5.  Almost all 
previous specific models of radiative Majorana neutrino masses are embodied 
in Fig.~3.  The detailed structures of this and the other two diagrams are 
systematically described here for the first time.  The new particles involved 
are possibly of order 100 GeV, in which case experimental verification 
is within reasonable reach in the near future.
\vskip 0.5in
\begin{center} {ACKNOWLEDGEMENT}
\end{center}

This work was supported in part by the U.~S.~Department of Energy under 
Grant No.~DE-FG03-94ER40837.  I thank K. S. Babu, R. Foot, A. S. Joshipura, 
H. Murayama, and J. Pantaleone for communications.

\bibliographystyle{unsrt}


\newpage
\begin{center}
\begin{picture}(160,80)(0,0)
\ArrowLine(20,40)(80,25)
\Text(50,45)[c]{$\nu_i$}
\DashArrowLine(140,40)(80,25)6
\Text(110,45)[c]{$\phi^0$}
\Line(80,25)(80,-25)
\Text(85,0)[l]{$N$}
\ArrowLine(20,-40)(80,-25)
\Text(50,-20)[c]{$\nu_j$}
\DashArrowLine(140,-40)(80,-25)6
\Text(110,-20)[c]{$\phi^0$}
\end{picture}
\vskip 0.8in
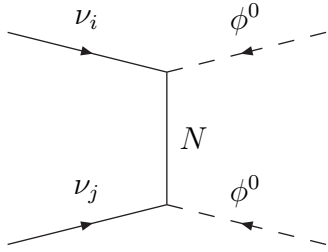
{\bf Fig.~1.} ~ Tree-level realization of the effective operator (3) with 
heavy fermion singlet.

\begin{picture}(160,80)(0,0)
\ArrowLine(0,40)(40,0)
\Text(30,30)[r]{$\nu_i$}
\ArrowLine(0,-40)(40,0)
\Text(30,-30)[r]{$\nu_j$}
\DashArrowLine(120,0)(40,0)6
\Text(80,10)[c]{$\xi^0$}
\DashArrowLine(160,40)(120,0)6
\Text(130,30)[l]{$\phi^0$}
\DashArrowLine(160,-40)(120,0)6
\Text(130,-30)[l]{$\phi^0$}
\end{picture}
\vskip 0.8in
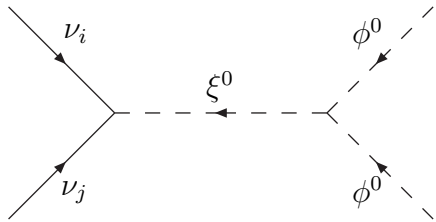
{\bf Fig.~2.} ~ Tree-level realization of the effective operator (3) with 
heavy scalar triplet.

\begin{picture}(200,120)(0,0)
\ArrowLine(20,0)(60,0)
\Text(40,-8)[c]{$\nu_i$}
\DashArrowArc(100,0)(40,90,180)6
\Text(59,28)[c]{$\chi$}
\ArrowLine(100,0)(60,0)
\Text(80,-8)[c]{$\omega$}
\DashArrowLine(100,-40)(100,0)6
\Text(100,-47)[c]{$\phi^0$}
\ArrowLine(100,0)(140,0)
\Text(120,-8)[c]{$\omega^c$}
\ArrowLine(180,0)(140,0)
\Text(160,-8)[c]{$\nu_j$}
\DashArrowArcn(100,0)(40,90,0)6
\Text(142,28)[c]{$\eta$}
\DashArrowLine(100,80)(100,40)6
\Text(100,87)[c]{$\phi^0$}
\end{picture}
\vskip 0.8in
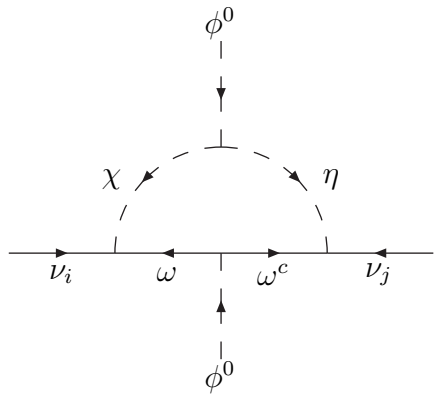
{\bf Fig.~3.} ~ First one-loop realization of the effective operator (3).
\end{center}

\newpage
\begin{center}
\begin{picture}(200,120)(0,0)
\ArrowLine(20,0)(60,0)
\ArrowLine(100,0)(60,0)
\ArrowLine(100,0)(140,0)
\ArrowLine(180,0)(140,0)
\DashArrowLine(68,72)(100,40)6
\DashArrowLine(132,72)(100,40)6
\DashArrowArc(100,0)(40,90,180)6
\DashArrowArcn(100,0)(40,90,0)6
\Text(40,-8)[c]{$\nu_i$}
\Text(80,-8)[c]{$\omega$}
\Text(120,-8)[c]{$\omega^c$}
\Text(160,-8)[c]{$\nu_j$}
\Text(59,28)[c]{$\chi$}
\Text(142,28)[c]{$\eta$}
\Text(60,80)[c]{$\phi^0$}
\Text(140,80)[c]{$\phi^0$}
\end{picture}
\vskip 0.5in
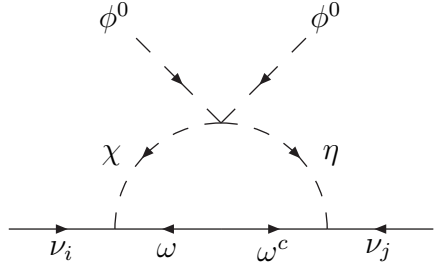
{\bf Fig.~4.} ~ Second one-loop realization of the effective operator (3).

\begin{picture}(200,80)(0,0)
\ArrowLine(-10,0)(30,0)
\ArrowLine(70,0)(30,0)
\ArrowLine(70,0)(100,0)
\ArrowLine(130,0)(100,0)
\ArrowLine(130,0)(170,0)
\ArrowLine(210,0)(170,0)
\DashArrowLine(70,-40)(70,0)6
\DashArrowLine(130,-40)(130,0)6
\DashArrowArcn(100,-41)(81,150,30)6
\Text(10,-8)[c]{$\nu_i$}
\Text(50,-8)[c]{$\omega$}
\Text(85,-8)[c]{$\sigma^c$}
\Text(115,-8)[c]{$\sigma$}
\Text(150,-8)[c]{$\omega^c$}
\Text(190,-8)[c]{$\nu_j$}
\Text(70,-47)[c]{$\phi^0$}
\Text(130,-47)[c]{$\phi^0$}
\Text(100,50)[c]{$\zeta$}
\end{picture}
\vskip 1.0in
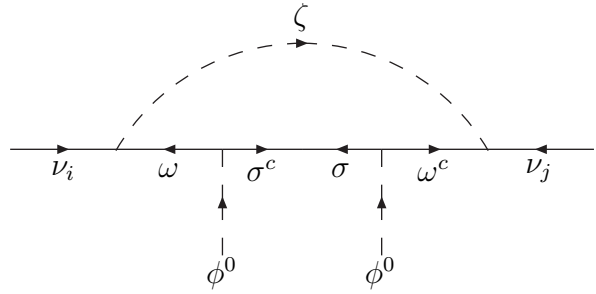
{\bf Fig.~5.} ~ Third one-loop realization of the effective operator (3).
\end{center}
\end{document}